\documentclass{aa}
\usepackage{psfig}

\newcommand{\ms}{\mbox{$M_{\sun}$}}

\newcommand{\rs}{\mbox{$R_{\sun}$}}
\def\aplt{\ {\raise-.5ex\hbox{$\buildrel<\over\sim$}}\ }
\def\apgt{\ {\raise-.5ex\hbox{$\buildrel>\over\sim$}}\ }
\newcommand{\mc}{\mbox {$M_{Ch}$}}
\newcommand{\md}{\mbox {$\dot{M}$}}

\begin{document}

\title{Is KPD 1930+2752 a good candidate Type Ia supernova progenitor?}

\author{E. Ergma\inst{1,2}, A. V. Fedorova\inst{3}
 and L. R. Yungelson \inst {3,4}} 

\offprints{Ene Ergma}

\institute{Physics Department, Tartu University, \"Ulikooli 18, 
      50090, Tartu, Estonia;~ene@physic.ut.ee
\and
 Observatory, P.O.Box 14, FIN-00014, University of Helsinki, Finland
\and
 Institute of Astronomy,
  48 Pyatnitskaya Str., 109017 Moscow, Russia;~afed,~lry@inasan.rssi.ru
\and
Astronomical Institute ``Anton Pannekoek'', 
        Kruislaan 403, NL-1098 SJ Amsterdam, the Netherlands}
\date{ Received \today}

\titlerunning{Is KPD~1930+2752 a good SN\,Ia progenitor?}
\authorrunning{E. Ergma et al.}

\abstract{We investigate the evolution of a binary system which initially
 has an orbital period of 2$^h$17$^m$\ and contains a 0.5\,\ms\ helium
 star with a white dwarf companion of 0.97\,\ms, similarly to suggested
 SN\,Ia candidate progenitor KPD 1930+2752. We show that the helium star completes
 core helium burning and becomes a white dwarf before components merge.
 The most probable outcome of the merger of components is formation of
 a massive white dwarf, despite initially the total mass of the system is
 above the Chandrasekhar mass. 
}

\maketitle

\section{Introduction}
The double-degenerate (DD) model for progenitors of type Ia supernovae
 (Tutukov \& Yungelson 1981; Webbink 1984; Iben \& Tutukov 1984)
 considers a binary  with the total mass of  white dwarf components
 higher than the Chandrasekhar limit, which   merges in less than Hubble
 time due to the loss of angular momentum via gravitational wave radiation.
 Type Ia supernova (SN Ia)  is supposed to result from an explosive carbon
 burning in the merger product. This model encounters two major problems:
 (i) none of the detected   DD systems with sufficiently short orbital
period has an estimated total mass of the components  above  the
 Chandrasekhar limit (see Maxted \& Marsh 1999);  (ii) it is not clear
 whether the carbon detonation may be initiated in the merger product.

 The first  of these problems has been attempted to be
 solved by systematic surveys for DD (e.g., the latest are by 
 Saffer et al. 1998;  Koester et al. 2001), which  still didn't give
 definite results. The second problem still awaits numerical solution (see, e.g., Segretain et al. 1997).

In meantime, helium-rich type B subdwarfs with white
 dwarf companions (sdB+wd    systems) have been suggested as candidate
 SNe\,Ia progenitors (Saffer et al. 1998; Marsh 2000; Maxted et al. 2000). In particular, the star KPD~1930+2752   
 (Downes 1986) has attracted attention. High speed photometry
 (Bill\'{e}res et al. 2000) and spectroscopy of the H$\alpha$ and
 HeI 6678\,\AA\ lines (Maxted et al. 2000) provide the evidence
 that this system is a binary with $P_{orb}=2^h17^m$.
 Maxted et al. show that the total mass of the binary is at least
 1.47$\pm 0.01\ms$, if  sdB star has a ``canonical'' mass of hot subdwarfs
 of 0.5\ms. This makes, to their opinion,   KPD~1930+2752  the first
 good candidate SN Ia progenitor. 

In this {\it Letter} we \textbf{discuss} the possible evolution and
 fate of a system similar to KPD~1930+2752. In Sec.\ref{sec:scen} we
 briefly consider formation and overall features of sdB+wd systems,
 in Sec.\ref{sec:code} some details of our evolutionary code are
 described. Numerical results are given in Sec.\ref{sec:res}. A discussion
 follows in Sec.\ref{sec:disc}. 

\section{Formation of low-mass helium star - white dwarf  systems}
\label{sec:scen}

One may start with an intermediate mass close
 binary: $M_1\approx 5 - 10\,\ms, M_2\approx 2.5 - 5.0\,\ms$. As a result
 of the Roche lobe overflow (RLOF) the primary component becomes a CO white
 dwarf. If the secondary experiences case B of RLOF, it becomes a $\sim
 (0.35 - 0.80)\,\ms$ helium star. In the core helium burning stage  it
 settles in the region of the $\log T_{eff} - \log g$\ diagram
 occupied by the stars which are spectroscopically classified as
 ``subdwarf B stars'', but in the evolutionary status nomenclature are
 called  ``extreme horizontal branch stars'' (EHB)\footnote{
 Maxted et al. (2001) specially notice  this terminological distinction; we follow the ``observational'' notation -- sdB.}.

\textbf{In an another scenario for formation of sdB + wd system the secondary component of a close binary may be a  $M_2 \aplt 2.5\,\ms$ star}. If the secondary fills the Roche lobe   when the mass of its helium core is still lower but sufficiently close to the helium ignition limit ($\sim 0.45\,\ms$ for the solar metallicity objects) the remnant of the secondary may  ignite He in the core but it will never become an AGB star (D'Cruz et al. 1996).    

Maxted et al. (2001) estimate that $69\pm9\%$ of all EHB stars are
 in short period binaries ($0.03 {\rm d}  \aplt P_{orb} \aplt 10 {\rm d}$).
Green et al. (2000) find that at least 2/3 of local disk sdB stars are binaries. Their survey suggests that most of them ``with periods of   
the order of hours or a few days have essentially invisible companions''. It's naturally to assume that these systems contain white dwarfs.  

However, only a few systems have confirmed white dwarf components. Both orbital period and masses of components are known  only
for KPD~1930+2752 and KPD~0422+5421 (Orosz \& Wade 1999) and only in  KPD~1930+2752  the total mass is close to \mc.

Helium stars with  $M \aplt 0.8$\,\ms\ do not expand in the core
helium burning stage (Paczy\'{n}ski 1971). In a wd+sdB system,  if the core helium burning time $t_{He} \sim (1 - 2)\,
10^8\, {\rm yr}$\ is shorter than the merger time due to 
 gravitational wave radiation
\begin{equation}
\label{eq:tgr} t_{\rm GR} \approx 1.5 \times  10^8 a^4 M_1^{-1} M_2^{-1}
 (M_1 + M_2)^{-1}~~ {\rm yr},
\end{equation}
where $a$ is the orbital separation (in \rs), $M_{1,2}$ - the masses of
 components (in \ms),  helium star may evolve directly into a
 CO white dwarf. In the case of KPD~1930+2752
 these two time scales,  $t_{He}$ and $t_{\rm GR}$, are comparable and the fate of the system has to
 be explored numerically.
 
\section{The evolutionary code}
\label{sec:code}

For the present study we \textbf{applied} an upgraded version of the evolutionary code which was used \textbf{before} for   
 the calculation of  evolution of binaries with helium secondaries (Ergma \& Fedorova  1990).

We have implemented the equation of state for the helium-rich matter  given by 
Saumon et al. (1995) and for carbon by Fontaine et al. (1977).
 Helium burning rate has been estimated according to Caughlan et al.
 (1985). Neutrino losses have been calculated after
 Beaudet et al. (1967). We have used opacity tables of
 Iglesias \& Rogers (1996) and Alexander \& Ferguson (1994). 
The mass loss rate in the Roche lobe filling stage has been calculated
 following Kolb \& Ritter (1990). The initial abundance of He in
 the model was  assumed to be $Y_c=0.98$, abundance of heavy elements 
 $Z= 0.02$.

\section{Results of calculations}
\label{sec:res} 

To understand the  fate of  KPD~1930+2752 
we have calculated the evolution of a 0.5\,\ms\
helium star through the core helium burning stage. 

\begin{figure}
\psfig{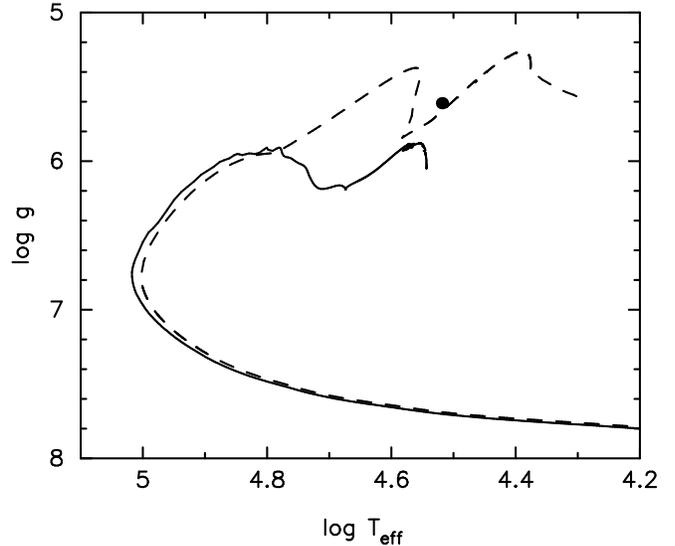}
\caption[]{Evolutionary track of a 0.5\,\ms\ helium star in the $T_{eff} - \log g$\
 plane. Solid line is for pure helium star, dashed line is for a 
 star which initially had 0.001\,\ms\ hydrogen envelope. Heavy dot
 marks position of KPD~1930+2752.}
\label{fig:g-t}
\end{figure}

For pure helium $0.5\,\ms$\ star the effective temperature and
 surface gravity  are higher 
than the measured $T_{eff}$=33\,000\,K and $\log g =5.61$ for
 KPD~1930+2752. Both $T_{eff}$ and $g$ are slightly lower if the  star has
 initially a low-mass hydrogen envelope ($<$ 0.001$M_\odot$), which
 is then lost during core helium burning phase
 (Fig.\ref{fig:g-t})\footnote{One has also to bear in mind that
 the mass of helium star in KPD 1930+2752 is {\it assigned} but not
 measured directly.}.

\begin{figure}[h]
\psfig{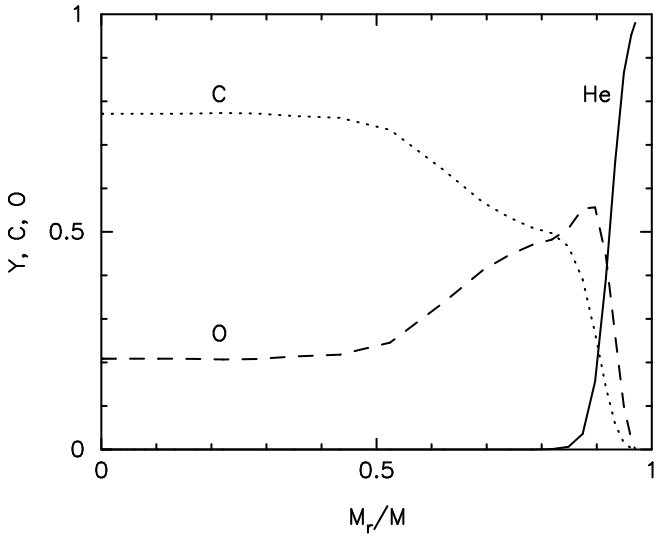}
\caption[]{Distribution of helium, carbon and oxygen in the nascent
 white dwarf formed by a 0.5\,\ms\ helium star.}
\label{fig:chem}
\end{figure}

The helium in the core of the model was exhausted in $\sim 1.5\times
 10^8$\,yr\ and a progenitor of a low mass carbon-oxygen white dwarf with
 a helium envelope was formed. The distribution of chemical species in the model upon completion of the helium  burning is shown in Fig.\,\ref{fig:chem}. 

It was assumed that the helium star has a 0.97\,$M_\odot$ companion and
the   initial $P_{orb}$ was set equal to 2$^h$17$^m$. Variation of
 separation of components was \textbf{then} followed assuming standard equation for the angular momentum loss via gravitational waves 
(Landau \& Lifshitz 1971).

During helium burning and subsequent cooling phases orbital period
  decreases due to gravitational wave radiation and at $P_{orb}
 \approx~1.^m5$\ the less massive white dwarf fills its Roche lobe.
 The following mass exchange stage is crucial for the  fate of the system.
 In other words,  this phase determines whether SN\,Ia occurs.
 Two scenarii are possible.

\begin{figure}[t]
\psfig{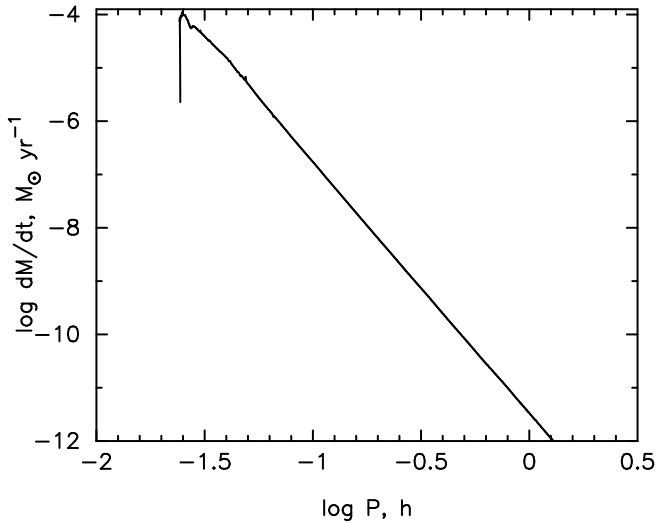}
\caption[]{Mass loss rate by the Roche-lobe filling white dwarf vs.
 orbital period of the binary.}
\label{fig:p-md}
\end{figure}

\subsection{Delayed merging scenario}
\label{sec:delayed} 

White dwarfs overfilling Roche lobe are subject to
a  dynamical instability if  the mass ratio of components exceeds a
 certain critical value $\sim 2/3$ (Pringle \& Webbink 1975; see also 
 Han \& Webbink (1999) and Nelemans et al. (2001) for the most recent
 discussion  of  stability and energetics of mass transfer in double
 white dwarfs).  Mass exchange in a $0.5\,\ms+0.97\,\ms$ binary white dwarf is expected
 to be dynamically stable, but $\dot{M}$\ has to exceed $\md_{Edd}$ for a $0.97\,\ms$ white dwarf.  We assumed that white
 dwarf accretes at the Eddington rate and that excess of the matter is lost
 from the system with the specific angular momentum equal to that of the
 accretor. This gives an additional stabilising effect to the mass
 exchange. The dependence of the mass loss rate on the orbital period \textbf{for  our model system} is shown in Fig.\ref{fig:p-md}.

After $\sim\,0.218$\,\ms\ is lost by the donor, mass accretion rate
 decreases below $\md_{Edd}$. At the end of the mass
 transfer 
$(t = 10^{10}$\, yr) orbital period has increased to $\sim 1.^h3$\
 and the mass of the secondary became less than 0.007\,\ms. Although
  primary mass has grown to 1.33\,\ms\ it remained below the limiting mass
 for the carbon burning. 

However, \textbf{the picture described above} almost certainly oversimplifies the
 actual evolution. First,  the liberated accretion energy has to be
 sufficient to evaporate the matter from the distance of  the order of
 the orbital separation of components. This requires $\md_2/\md_{Edd}
 \sim 10$, a condition which may be not fulfilled (Fig.\,\ref{fig:p-md}).
\textbf{If this is the case,} one may expect formation of a common  envelope and the merger of
  components due to dissipative orbital energy losses. An R CrB type star may be formed in this way (Iben et al. 1996). An another complication
 is brought in  by the presence of a He-rich layer atop the donor
 star (Fig.\,\ref{fig:chem}). 

 As our test calculations of accretion onto a white dwarf show, 
during the stage of accretion with $\md_{Edd} \sim 1.5\times 
10^{-5}\,M_\odot\,{\rm yr^{-1}}$ white dwarf accumulated a  helium envelope of
 $\sim 0.01$\,\ms\ and when the density at the bottom of the 
accreted layer  attained $\sim 2\times 10^5\,{\rm g\,cm^{-3}}$ helium
 burning started and a thermal flash developed.  We have terminated  our
 calculations when the temperature at the bottom of the accreted envelope 
 rose from $\sim 10^7$ to  $\sim 2.7\times 10^8$\,K. Expansion of the outer
 layers will lead to the common envelope formation and the merger of
 companions. \textbf{Some mass may be lost in this process.}

\subsection{Prompt merging scenario}
\label{sec:prompt}
It may well happen that in the real system the coalescence of white dwarfs
 occurs on a dynamical time scale (e.g., if the mass of the sdB star is
 higher than assumed, but still $t_{He} < t_{GR}$). Segretain et al. 
 (1997) presented a three-dimensional SPH
 simulation of the coalescence of  carbon--oxygen white
 dwarfs in a binary rather similar to the system expected to be formed
 by KPD~1930+2752: $M_1$=0.9$M_\odot$ and $M_2$=0.6$M_\odot$. The less
 massive white dwarf is disrupted on a dynamical time scale and becomes a
 thick disk around the more massive primary. Carbon ignition is likely to
 occur at the core--disk boundary, the hottest part of the merged
 configuration.  Since this region is only weakly degenerate, carbon
 ignition is expected to be non-violent and nuclear burning \textbf{will 
 propagate} inward, forming  an ONeMg core (Nomoto \& Iben 1985).
 Carbon continues to accrete due to viscous transport of momentum. 
 Generation  of energy by accretion is expected to transform
 the disk into a quasi-spherical envelope. A stationary configuration
 may form in which   carbon burns at the same rate as it is accreted by the
 core (Kawai et al. 1988). During this high luminosity phase mass
 loss through a stellar wind should take place. A significant part of the
 envelope may be lost and  a massive single white dwarf will be formed. 

Evolution of  KPD~1930+2752 may be different in some features, important
 for the fate of the system.  The disrupted less massive white dwarf will
 have rather thick helium surface layer ($\sim$ 0.04$M_\odot$) which  will
 be accreted by the companion the first.  Like in the case of carbon accretion,
 these external parts of the disrupted donor are heated by shocks and become
 the hottest part of the merger product: $T \sim (7-8)\times10^8\,{\rm K}$. 
 The time scale for the complete conversion of helium into carbon
 (neglecting all reactions other than the $3\alpha$-reaction) is
 \begin{equation}
\tau_{3\alpha} \sim Y^{-3}T_8^3\rho_6^{-2}\exp(-14.33+43.2/T_8)~~{\rm s},
\end{equation}
where $Y$\ is the abundance of ${\rm He^4}$ by mass, $T_8$ and $\rho_6$ are
 the temperature and the density in the units of $10^8$\,K and $10^6\,{\rm  g\
cm^{-3}}$, respectively (Iben \& Tutukov 1991). For $Y=1$, $\rho \sim {\rm 10^4-10^5\, g\,cm^{-3}}$\
 and $T_8\sim 7$\ one has  $\tau_{3\alpha} \sim {\rm 10\,s}$\
 for $\rho_6=0.1$\ or 1000\,s for $\rho_6=0.01$. The energy produced by
 the burning of 0.04$M_\odot$\ of helium is $\sim 10^{50}$\, erg. This
 energy is comparable to the binding energy of the whole envelope
 (former less massive dwarf). Very fast release of the energy in a weakly
 degenerate matter may result in expansion and loss of the envelope. 
On the other hand, carbon burning may start and propagate inward. As a
 result, like in the case of {\it delayed} merging, one may expect a
 formation of an ONeMg white dwarf, but of relatively lower mass
 ($\aplt 1\,\ms$). However, these inferences, for both cases of
 {\it delayed} and {\it prompt} merging, have to be verified by
 hydrodynamic calculations.

\section{Discussion and conclusion}
\label{sec:disc}
In the absence of observed candidate double degenerate progenitors of SNe\,Ia,
 systems containing a white dwarf with a low mass helium companion and
 merging due to the loss of angular momentum via GWR were suggested as 
 SNe\,Ia progenitors.  Two routes to explosion may be envisioned. 

 First,
 helium star may fill its Roche lobe while still burning He in the core.
 If He-burning isn't much advanced, mass exchange rate upon RLOF is
 expected to be $\sim 3 \times 10^{-8}\,M_\odot\,{\rm yr^{-1}}$ (Savonije    et al. 1986; Tutukov \& Fedorova 1989; Ergma \& Fedorova 1990)
 and this may result in the so called ``edge-lit'' detonation   after
 accretion of $\sim 0.1\ms$\ of He (Livne 1990). 
An another possibility is
 the merger of components  after helium star also becomes a white dwarf.
 Thus, the outcome of evolution depends on the relation of the time scales
 of GWR and core helium burning. We have shown, that for a system similar
 to KPD~1930+2752 the merger is more probable, given the state of the art  input
 physics of our evolutionary code. 

However, the situation differs from
 the ``standard'' picture of the merger of two CO white dwarfs with a
 dynamical disruption of the less massive dwarf,  due to presence of
 a helium mantle ($\sim 0.04\,\ms$) atop the latter. If mass transfer to  the more massive dwarf is initially stable, accreted  He layer may experience a thermal flash, 
resulting in its expansion, formation of a common envelope and the merger of
 two cores, accompanied by some mass loss. Another possibility is a dynamical
 merger, in which He will be ignited at the core-envelope interface of
 the merger product; about $\sim 10^{50}$\,erg may be released then in
 10 - 1000 s and  expulsion of most of the material of the disrupted
 dwarf may be expected.       In both cases it's awaited that 
 carbon will start burning in the outer layers of the core of the merger
 product and the flame will propagate inward, forming an ONeMg white
 dwarf. Thus we conclude that  KPD~1930+2752 will not produce a SN\,Ia.
 Three-dimensional hydrodynamical calculations of the merging process
 including nuclear burning are necessary to verify these speculations.
  
\begin{acknowledgements} LRY
 acknowledges warm hospitality  of the Astronomical
 Institute ``Anton Pannekoek'' and  support from NOVA and NWO Spinoza grant to E. P. J. van den Heuvel. EE acknowledges warm hospitality of the 
Astronomical Observatory of the Helsinki University. This work was partially
supported by the grant N 157992 from the Academy of Finland to  Dr. O. Vilhu,
 ESF grant N 4338, RFBR grant 99-02-16037 and Program ``Astronomy'' grant 1.4.4.1. 
\end{acknowledgements}


\begin{thebibliography}{}
\bibitem{}Alexander, D. R., \& Ferguson, J. W. 1994, ApJ, 437, 879

\bibitem{}Caughlan, G. R., Fowler, W. A., Harris, M. J., \& Zimmerman,
 B. A. 1985, Atom. Data and Nucl. Data Tables, 32, 197

\bibitem{}Beaudet, G., Petrosian, V., \& Salpeter, E. E. 1967, ApJ, 150, 979

\bibitem{}Billeres, M., Fontaine, G., Brassard, P., Charpinet, S., Liebert,
 J., et al. 2000, ApJ, 530, 441
  
\bibitem{}D'Cruz, N. L., Dorman, B., Rood, R. T., \& O'Connell, R. W.
1996, ApJ, 466, 359
    
\bibitem{}Downes, R.A. 1986, ApJS, 61, 569

\bibitem{}Ergma, E., \& Fedorova, A. 1990, Ap\&SS, 163, 142

\bibitem{}Fontaine, G., Graboske, H. C., \& van Horn, H. M. 1977, ApJS, 35, 293

\bibitem{}Green, E. M., Liebert, J., \& Saffer, R. A. 2000, BAAS, 197, 4601
    
\bibitem{}Han, Z., \& Webbink, R. F. 1999, A\&A, 349, L17

\bibitem{}Iben, I.\,Jr., \& Tutukov, A. V. 1984, ApJS, 54, 355

\bibitem{}Iben, I.\,Jr., \& Tutukov, A. V. 1991, ApJ, 370, 615

\bibitem{}Iben, I.\,Jr., Tutukov, A. V., \& Yungelson, L. R.
1996, ApJ, 456, 750

\bibitem{}Iglesias, C. A., \& Rogers, F. J. 1996, ApJ, 464, 943

\bibitem{}Kawai, Y., Saio, H., \& Nomoto, K. 1988, ApJ, 328, 207

\bibitem{}Koester, D., Napiwotzki, R., Chriestlieb, N. et al. 2001,
 A\&A, submitted

\bibitem{}Kolb,  U., \& Ritter, H. 1990, A\&A, 236, 385

\bibitem{} Landau, L. D. \& Lifshitz, E. M. 1971, Classical theory of
 fields (Oxford: Pergamon)

\bibitem{} Livne, E. 1990, ApJ, 354, L53.

\bibitem{}Marsh, T. R. 2000, New Astr. Rev., 44, 119

\bibitem{}Maxted, P. F. L. \&  Marsh, T. R. 1999, MNRAS, 307, 122   

\bibitem{}Maxted, P. F. L., Marsh, T. R., \& North, R. C. 2000, MNRAS, 317, L41

\bibitem{}Maxted, P. F. L., Heber, U., Marsh, T. R., \& North, R. C. 2001,
 astro-ph/0103342

\bibitem{}Nelemans, G., Portegies Zwart, S. F., Verbunt, F., \& Yungelson,
 L. R. 2001, A\&A, 368, 939

\bibitem{}Nomoto, K. \& Iben, I.\,Jr. 1985, ApJ, 297, 531

\bibitem{}Orosz, J. A. \& Wade, R. A. 1999, MNRAS, 310, 773   

\bibitem{}Paczy\'{n}ski, B. 1971, Acta Astr., 21, 1

\bibitem{}Pringle, J. E., \& Webbink, R. F. 1975, MNRAS, 172, 493

\bibitem{}Saffer, R. A., Livio, M. \& Yungelson,  L. R. 1998, ApJ, 502, 394

\bibitem{}Saumon, D., Chabrier, G., \& van Horn, H. M. 1995, ApJS, 99, 713

\bibitem{} Savonije, G. J., de Kool, M., \& van den Heuvel, E. P. J. 1986,
 A\&A, 155, 51

\bibitem{}Segretain, L., Chabrier, G., Mochkovitch, R. 1997, ApJ, 481, 355

\bibitem{}Tutukov, A. V., \& Fedorova, A. V. 1989, SvA, 33, 606

\bibitem{}Tutukov, A. V., \& Yungelson, L. R. 1981, Nauchn. Inform., 49, 3

\bibitem{}Webbink, R. F. 1984, ApJ, 227, 355.

\end{thebibliography}
\end{document}